\documentclass{ws-ijmpc}

\begin{document}

\markboth{Nuno Crokidakis}
{Modeling the early evolution of the COVID-19 in Brazil: results from a Susceptible-Infectious-Quarantined-Recovered (SIQR) model}


\catchline{}{}{}{}{}

\title{Modeling the early evolution of the COVID-19 in Brazil: results from a Susceptible-Infectious-Quarantined-Recovered (SIQR) model}

\author{Nuno Crokidakis}

\address{Instituto de F\'isica, Universidade Federal Fluminense \\
Niter\'oi/RJ, Brazil\\
nuno@mail.if.uff.br}

\maketitle

\begin{history}
\received{Day Month Year}
\revised{Day Month Year}
\end{history}

\begin{abstract}
The world evolution of the Severe acute respiratory syndrome coronavirus 2 (SARS-Cov2 or simply COVID-19) led the World Health Organization to declare it a pandemic. The disease appeared in China in December 2019, and it has spread fast around the world, specially in european countries like Italy and Spain. The first reported case in Brazil was recorded in February 26, and after that the number of cases growed fast. In order to slow down the initial growth of the disease through the country, confirmed positive cases were isolated to not transmit the disease. To better understand the early evolution of COVID-19 in Brazil, we apply a Susceptible-Infectious-Quarantined-Recovered (SIQR) model to the analysis of data from the Brazilian Department of Health, obtained from February 26, 2020 through March 25, 2020. Based on analyical and numerical results, as well on the data, the basic reproduction number is estimated to $R_{0}=5.25$. In addition, we estimate that the ratio unidentified infectious individuals and confirmed cases at the beginning of the epidemic is about $10$, in agreement with previous studies. We also estimated the epidemic doubling time to be $2.72$ days.

\keywords{Dynamics of social systems, Collective phenomena, Data Analysis}

\end{abstract}

\ccode{PACS Nos.: 87.23.Ge, 89.20.-a, 89.65.-s, 89.75.Fb}

\section{Introduction}

The evolution of epidemics is one of the most dangerous problems for a society. The humanity faced severe pandemics during its evolution, like the Spanish flu in 1917, the Honk Kong flu (H3N2) of 1968 and the swine flu (H1N1) in 2009. Several efforts were done since 70's in order to understand the mathematical evolution and spreading of diseases \cite{anderson,bailey}.

Recently, in  December, 2019, Wuhan  city,  the capital of Hubei province in China,  became the centre of an outbreak of pneumonia of unknown cause. By January  7,  2020,  Chinese  scientists  had  isolated  a  novel  coronavirus,  severe  acute  respiratory syndrome coronavirus 2 (SARS-CoV-2), from these patients with virus-infected pneumonia \cite{phelan,baker},  which  was  later  designated  coronavirus disease 2019 (COVID-19) in February, 2020, by World Health Organization.

In order to better understand such new disease, a lot of papers and preprints were published in the last months \cite{li_science,biswas,bin,pedersen,zhou,botha,liu,kraemer,zullo,cavanagh,ferguson,lai,zhao,gayle,muniz-rodriguez,yin,gonzalez,roques,chang,faggian}, analyzing various properties of the COVID-19, as well as modeling its evolution in many countries. Our target in this work is to performe a data analysis of the evolution of COVID-19 in Brazil. We collected data from the Brazilian Department of Health, and apply a Susceptible-Infectious-Quarantined-Recovered (SIQR) model to analyze the COVID-19 dynamics.

This work is organized as follows. In Section II we present the SIQR model, and define its parameters. In Section 3 we perform the data analysis and present analytical and numerical results based on the SIQR model. Finally, we present a discussion in Section IV. Some numerical details are presented in an Appendix.


\section{Model}

The model considered in this work is a variant of the Susceptible-Infected-Recovered (SIR) model \cite{anderson,bailey}. In addition to the usual compartments Susceptible (S), Infected (I) and Recovered (R), an extra compartment is considered, namely Quarantined (Q) \cite{hethcote,pedersen,bin}. $N$ is the total number of individuals in the population, assumed constant since we are studying the early phase of the epidemic. In this case, we have the normalization condition at each time step, i.e., $N(t)=S(t)+I(t)+Q(t)+R(t)$.

Notice that both I and Q individuals are infected, but as discussed in \cite{bin} this separation is convenient because it models the fact that many governments (including the Brazilian one) are forcing individuals tested positive (confirmed cases) to self-isolate from the community, and also because it distinguishes between the infectious people who do self-isolate, and/or those who do not (mostly likely because they have not developed the symptoms of the disease and are not aware of actually being infectious). Thus, the SIQR model may be described by the following rate equations:
\begin{eqnarray} \label{eq1}
\frac{dS}{dt} & = & -\beta\,S\,I/N  \\ \label{eq2}
\frac{dI}{dt} & = & \beta\,S\,I/N - (\alpha+\eta)\,I \\ \label{eq3}
\frac{dQ}{dt} & = & \eta\,I - \gamma\,Q \\ \label{eq4}
\frac{dR}{dt} & = & \gamma\,Q + \alpha\,I
\end{eqnarray}

In the above equations, $\beta$ denotes the infection rate, $\alpha$ is a rate that quantifies the recovering of asymptomatic individuals and $\eta$ is the rate of detection of new cases \footnote{In this case, an infected individual is positive tested and becomes a confirmed case. This individual is isolated (quarantined).}. Finally, $\gamma$ stands for the recovering of quarantined individuals.


\section{Results}

Looking at the data \cite{ministerio}, the first case in Brazil was reported in February 26. Thus, at the present moment a relatively small fraction of the Brazilian population has been found positive for COVID-19, which means we are still in the early phase of the epidemic where we have $S/N \approx 1$. In this case Eq. \eqref{eq2} can be approximated to
\begin{equation} \label{eq5}
\frac{dI}{dt} = [\beta-(\alpha+\eta)]\,I ~,
\end{equation}
\noindent
that can be directed integrated to obtain
\begin{equation} \label{eq6}
I(t)=I_{0}\,e^{[\beta-(\alpha+\eta)]\,t}   ~, 
\end{equation}
\noindent
where $I_{0}$ is the number of infectious individuals at the beginning of the outbreak. Eq. \eqref{eq6} can be rewriten as $I(t)=I_{0}\,e^{(\alpha+\eta)(R_{0}-1)\,t}$, and one can obtain the expression for the basic reproduction number $R_{0}$,
\begin{equation} \label{eq7}
R_{0} = \frac{\beta}{\alpha+\eta} ~.
\end{equation}
\noindent
As it is well known, the basic reproduction number $R_{0}$ is an indicator of the occurrence of an outbreak (if $R_{0}>1$). We will see in the following that one can estimate its value from the data.

As discussed in \cite{pedersen}, the number of individuals that have been confirmed positive for COVID-19 and put in isolation does not correspond to $I$ but to $Q+R$. One can found a relevant analytical expression summing Eqs. \eqref{eq3} and \eqref{eq4},
\begin{equation} \label{eq8}
\frac{d\,[Q+R](t)}{dt} = (\alpha+\eta)\,I(t) ~.
\end{equation}

\begin{figure}[t]
\begin{center}
\vspace{3mm}
\includegraphics[width=0.6\textwidth,angle=0]{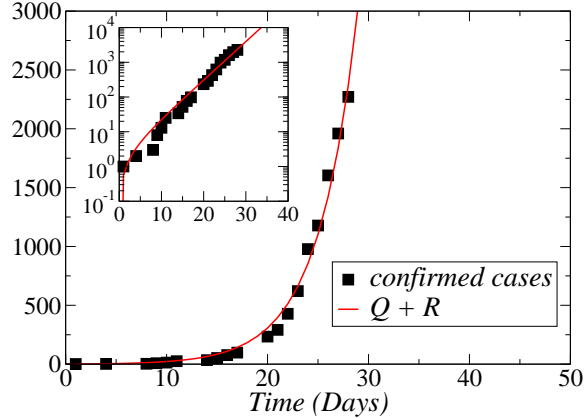}
\end{center}
\caption{(Color online) Comparison between the number of confirmed cases in Brazil (squares) and Eq. \eqref{eq9} (full line). The parameters are $I_{0}=8, \beta=0.315, \alpha+\eta=0.06$, as discussed in the text. In the inset we exhibit the graphic in the log-linear scale. Data was collected from February 26, 2020 through March 25, 2020.}
\label{fig1}
\end{figure}

Substituting the result \eqref{eq6} in Eq. \eqref{eq8} and integrating over $t$, one obtains
\begin{equation} \label{eq9}
[Q+R](t) = \frac{\alpha+\eta}{\beta-(\alpha+\eta)}\,I_{0}\,(e^{[\beta-(\alpha+\eta)]\,t}-1) ~.
\end{equation}

We fitted Eq. \eqref{eq9} to the the Brazilian COVID-19 data \cite{ministerio} from February 26, 2020 through March 25, 2020. The estimated values were $(\alpha+\eta)\,I_{0}=0.482$ and $\beta-(\alpha+\eta)=0.255$ (see details in the Appendix). Based on data, we take $I_{0}=8$ (see Appendix), which gives us $\alpha+\eta=0.06$ and $\beta=0.315$. Considering those estimates, we plot in Fig. \ref{fig1} the temporal evolution of the number of cases together with Eq. \eqref{eq9}.

Based on the above-mentioned fitted values, one can estimate from Eq. \eqref{eq7} $R_{0}=5.25$, which is in line with previous estimates of $R_{0}$ falling between 1.4 and 6.5 \cite{li_science,pedersen,liu,lai,gayle}. In addition, one can estimate the epidemic doubling time, that characterize the sequence of intervals at which the cumulative incidence doubles. From the above result $I(t)=I_{0}\,e^{(\alpha+\eta)(R_{0}-1)\,t}$ one can obtain the doubling time as $\tau=ln\,2/[(\alpha+\eta)\,(R_{0}-1)]$. Based on the above estimated parameters, we have $\tau=2.72$ days, which falls in the range $1.4 < \tau < 3.0$ estimated in China \cite{muniz-rodriguez}.

\begin{figure}[t]
\begin{center}
\vspace{3mm}
\includegraphics[width=0.6\textwidth,angle=0]{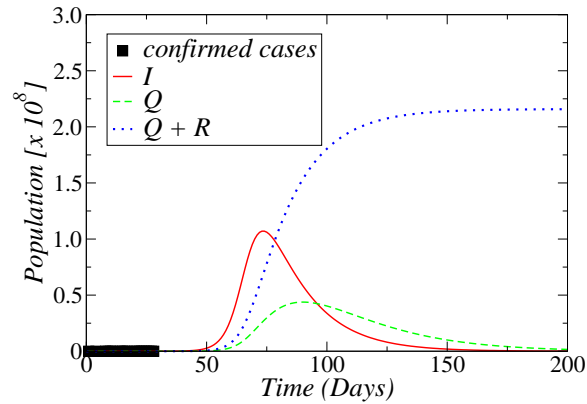}
\end{center}
\caption{(Color online) Time evolution of the number of Infected (I), Quarantined (Q) and total confirmed cases (Q+R), obtained by the numerical integration of the Eqs. \eqref{eq1} to \eqref{eq4}. Data of confirmed cases are exhibited as well (squares). The parameters are $I_{0}=8, \beta=0.315, \eta=\alpha=0.03$ and $\gamma=0.04$.}
\label{fig2}
\end{figure}

As discussed in the Appendix, we estimate $\eta=0.03$, $\alpha=0.03$ and $\gamma=0.04$. Considering those values, as well as the previous estimated values of $\beta$ and $I_{0}$, we plot in Fig. \ref{fig2} the time evolution of the number of Infected (I), Quarantined (Q) and total confirmed cases (Q+R), obtained by the numerical integration of the Eqs. \eqref{eq1} to \eqref{eq4}. For these curves, we considered $N$ as the total brazilian population, $N=2.17$ x $10^{8}$. One can see that the number of infected and nonconfirmed cases $I$ grows faster than the number of confirmed and isolated individuals Q. This unbalance is observed in all the world, since there is a huge number of undocumented infection cases for the COVID-19, as discussed in a recent work \cite{li_science}.

For better quantify the unbalance among unidentified infectious individuals and confirmed and isolated cases, one can discuss about the ratio $I/Q$. Taking the ratio of Eqs. \eqref{eq3} and \eqref{eq2}, and considering the approximated result for short times, Eq. \eqref{eq5}, one can obtain
\begin{equation}\label{eq10}
\frac{I}{Q} = \frac{\gamma+\beta-(\alpha+\eta)}{\eta}  ~.
\end{equation}
\noindent
Based on the estimated values of the parameters, Eq. \eqref{eq10} gives us $I/Q \approx 9.83$, i.e., for each patient in quarantine approximately \textit{ten} infectious individuals are present in the population \footnote{This number of course depends on a precise estimate of the value of $\gamma$, but usually the recovered rates are small, in the range $0.01 < \gamma < 0.05$, which give us $8.83 < (I/Q) < 10.16$.}. This is in agreement with a recent work that states that $86\%$ of all infections were undocumented in China \cite{li_science}. Thus, there are still many unidentified cases that do not appear in the official statistics.


\section{Discussion}   

The number of confirmed cases of COVID-19 in Brazil is growing exponentially fast. Based on the data, we considered a Susceptible-Infectious-Quarantined-Recovered (SIQR) on a fully-connected population. Despite the simplicity of the model, we can made estimates of the infection rate, the rate which individuals are isolated from the population and some others, as well as the basic reproduction number and the doubling time of the epidemics, based on the data available at the site of the Brazilian Department of Health.

For the considered parameters, we observed that the number of quarantined individuals grows fast (exponentially), stabilizes and after it decays to zero, as it is standard in compartmental models. Based on the data, we can see that the number of such isolated individuals grows until Day 90 from the beginning of the disesase spreading (February 26, 2020). Thus, the model predicts that the maximum number of isolated individuals will occur about May 25, 2020. This is in line with a recent estimate \cite{ferguson}. This peak is associated with the isolation of about $20 \%$ of the brazilian population.

This pessimistic estimate can be modified if the government imposes limitations for the population, as was done in some countries like Italy, Spain and India. In order to reduce the growing of the number of cases, social isolation was suggested in some brazilian states, but it was only a suggestion by the local governments, i.e., there is no mandatory quarantine. In a recent work \cite{ferguson}, the authors analyzed the potential role of non-pharmaceutical interventions in UK and USA. They conclude that the effectiveness of any one intervention in isolations likely to  be  limited, requiring multiple interventions to be combined to have a substantial impact on transmission. 

Other recent work \cite{kraemer} studied the role of people mobility in the diffusion of the COVID-19 in China. They show that travel restrictions are useful in the early stage of an outbreak when it is confined to a certain area that acts as a major source. In the case of Brazil, the major sources are the cities of Sao Paulo and Rio de Janeiro, and immediate actions of restrictions in mobility need to be adopted to control the spread of COVID-19 \cite{nota}.


\section*{Acknowledgments}

The author acknowledges financial support from the Brazilian scientific funding agencies Conselho Nacional de Desenvolvimento Cient\'ifico e Tecnol\'ogico (CNPq) and Funda\c{c}\~ao de Amparo \`a Pesquisa do Estado do Rio de Janeiro (FAPERJ).


\appendix

\section{}

In this appendix we discuss about the estimates of the model's parameters. Considering Eq. \eqref{eq9}, we did a least squares fitting of the data considering a function $f(t)=(a/b)\,(e^{b\,t}-1)$. In comparison with Eq. \eqref{eq9}, we have the identification $a = (\alpha+\eta)\,I_{0}$ and $b = \beta - (\alpha+\eta)$. The fitting procedure gives us $a=0.482 \pm 0.125$ and $b = 0.255 \pm 0.012$. Thus, we have $(\alpha+\eta)\,I_{0}=0.482$ and $\beta - (\alpha+\eta)=0.255$.

The time in our graphics were counted after the first confirmed case in Brazil, i.e., Day 1 was February 26, 2020. The number of confirmed cases kept almost constant for some days, and start to grow faster since day 9 (March 5, 2020), where 8 cases were confirmed. In this case, we take the number of initial cases as $I_{0}=8$ for the model's purposes. Thus, from $(\alpha+\eta)\,I_{0}=0.482$  we obtained $(\alpha+\eta)=0.060 \pm 0.002$. Considering this result, from $\beta - (\alpha+\eta)=0.255$ we obtained $\beta=0.315 \pm 0.010$.

As discussed in \cite{pedersen}, the parameter $\eta$ is related to the time until patients are tested positive and isolated, but also to the fraction of all infectious individuals that are tested positive. These are mostly symptomatic patients, which we assume are isolated soon after the incubation time is over and first symptoms appear. The incubation time means the time between catching the virus and beginning to have symptoms of the disease. Most estimates of the incubation period for COVID-19 range from 1-14 days, most commonly around $\approx 5$ days \cite{incubation_times}. Letting $\delta$ denote the fraction of infectious individuals entering Q, we obtain $\eta = \delta$ x $0.2$.

It was reported that $\approx 50\%$ of the population is asymptomatic \cite{japao_infect}, but some milder cases may also go unnoticed and not end in isolation. Considering the very small quantity of tests made in brazilians, we assume that $\delta = 15\%$ of infectious individuals are tested after an average of 5 days. We thus set $\eta = 0.15$ x $0.2 = 0.03$. From the above obtained relation, $(\alpha+\eta)=0.06$, we have $\alpha=0.03\, (\pm 0.003)$.

Finally, until March 25, 2020, the number of recovered individuals in Brazil is very small, and it is hard to obtain an estimate of $\gamma$. In this case, for most results we considered $\gamma=0.04$, in line with Ref. \cite{pedersen}. Notice that the main result of the model, the total number of confirmed cases $Q+R$, do not depend on $\gamma$, as we can see in Eq. \eqref{eq9}.


\end{document}